\documentclass[fleqn,usenatbib]{mnras}

\usepackage{newtxtext,newtxmath}
\usepackage[T1]{fontenc}
\usepackage{soul}

\DeclareRobustCommand{\VAN}[3]{#2}
\let\VANthebibliography\thebibliography
\def\thebibliography{\DeclareRobustCommand{\VAN}[3]{##3}\VANthebibliography}

\usepackage{graphicx}	% Including figure files
\usepackage{amsmath}	% Advanced maths commands
\usepackage{dcolumn}    % Align table columns on decimal point
\usepackage{bm}
\usepackage{xcolor}
\usepackage{ulem}

\title{Mass reconstruction and noise reduction with cosmic-web environments}

\author[Feng Fang et al.]{
Feng Fang$^{1}$
Yan-Chuan Cai$^{2}$\thanks{cai@roe.ac.uk}
Zhuoyang Li$^{1,3}$
Shiyu Yue$^{1}$
Weishan Zhu$^{1}$
and Longlong Feng$^{1}$\thanks{flonglong@mail.sysu.edu.cn}
\\
$^{1}$School of Physics and Astronomy, Sun Yat-Sen University, Zhuhai 519082, China \\
$^{2}$Institute for Astronomy, University of Edinburgh, Blackford Hill, Edinburgh, EH9 3HJ, UK \\
$^{3}$Department of Astronomy, Tsinghua University, Beijing 100084, China
}

\date{Accepted XXX. Received YYY; in original form ZZZ}

\pubyear{2023}

\begin{document}
\label{firstpage}
\pagerange{\pageref{firstpage}--\pageref{lastpage}}

\maketitle

\begin{abstract}
The clustering of galaxies and their connections to their initial conditions is a major means by which we learn about cosmology. However, the stochasticity between galaxies and their underlying matter field is a major limitation for precise measurements of galaxy clustering. Efforts have been made with an optimal weighting scheme to reduce this stochasticity using the mass-dependent clustering of dark matter haloes. Here, we show that this is not optimal. We demonstrate that the cosmic-web environments (voids, sheets, filaments \& knots) of haloes, when combined linearly with the linear bias, provide extra information for reducing stochasticity in terms of two-point statistics. Using the environmental information alone can increase the signal-to-noise of clustering by a factor of 3 better than the white-noise level at the scales of the baryon acoustic oscillations. The information about the environment and halo mass are complementary. Their combination increases the signal-to-noise by another factor of 2-3. The information about the cosmic web correlates with other properties of haloes, including halo concentrations and tidal forces -- all are related to the assembly bias of haloes. 
%For two-point statistics, these results are in qualitative agreement with what was presented in \citep{Kitaura2022} where higher order and non-local biases were included.
\end{abstract}

\begin{keywords}
large-scale structure of Universe -- dark matter -- Galaxy: halo -- methods: numerical
\end{keywords}

\section{Introduction}\label{sec:intro}
A major challenge in cosmology with large-scale structure is to reconstruct the initial conditions from observations of the late-time Universe. While the initial matter density field is thought to be linear and continuous, what we observe in the late-time Universe is a discrete sampling of a non-linear and non-Gaussian matter field, typically traced out by galaxies. The accuracy of the reconstruction determines the amount of cosmological information we can extract from observations. In terms of two-point statistics, which is the focus of our study, it can be quantified with the correlation coefficient $r$ between the galaxy number density field $\delta_g$ and the matter density field $\delta_m$,
%\FF{we can modify the main.tex without manually highlight the changes? with command line: "latexdiff old.tex new.tex > diff.tex", compiling this diff.tex will generate the PDF with changes highlighted}
\begin{equation}
\label{eq:cross}
r \equiv \frac{\langle\delta_g \delta_m\rangle}{\sqrt{\langle\delta_g^2\rangle \langle\delta_m^2\rangle}}.    
\end{equation}
The $\langle\,\rangle$ symbol denotes the ensemble average. To the first order, galaxies are linearly biased against matter. With the presence of shot noise $\epsilon$, we have $\delta_g=b\delta_m + \epsilon$, where $b$ is the linear bias of galaxies. Assuming that $\langle\epsilon \delta_m \rangle=0$, i.e., the noise is uncorrelated with the signal, we have $P_g=\langle\delta_g^2\rangle=b^2P_m+\langle\epsilon^2\rangle$, $P_{gm}=\langle \delta_g\delta_m\rangle=bP_m$, where $P_m$ is the matter power spectrum; and so 
\begin{equation}
\label{eq:S/N}
\frac{\langle\epsilon^2\rangle}{b^2P_m}=\frac{1-r^2}{r^2}.    
\end{equation}
All the above quantities are functions of the Fourier number $k$.
We can see that the presence of the noise, or stochasticity, degrades the correlation coefficient $r$ \citep{Dekel1999,Seljak2004, Cai2011,Lavaux&Jasche2021,Lee2022}. For two-point statistics, minimising stochasticity, increasing correlation coefficients and optimising mass reconstruction are different sides
of the same coin \footnote{Note that following the definitions, the reasoning above applies to two-point statistics only. It is not necessarily true for higher-order statistics.}. Reducing $\epsilon$ has the following observational implications: 

(1) increasing the signal-to-noise ratio for the clustering of galaxies \citep[e.g.][]{Seljak2009, Hamaus2010, Cai2011,Chan2012,Hamaus2012,Lee2022}, which is particularly beneficial for analyses of the baryon acoustic oscillations (BAO) with sparse tracers \citep[e.g.][]{Seo2007,Angulo2008, White2010, Beutler2014, Kitaura2014, Kitaura2016, Cohn2016, Lepori2017, Ding2018,Patej2018,zhao2019,Seo2022};

(2) increasing the correlations between the galaxy field and the matter field, allowing more cosmological information to be retrieved from the measurement of galaxy-galaxy lensing and reconstruction \citep[e.g.][]{Bonoli2009,Cai2012, Yang2015, Eriksen2018, Zhou2023b, Ma2023}.

$\epsilon$ was usually assumed to be white i.e., having no correlations among different frequencies and following a flat power spectrum of $1/\bar{n}$, with $\bar{n}$ being the number density of tracers. Recent work has shown that we can improve beyond this limit \citep{Bonoli2009, Seljak2009, Hamaus2010, Cai2011, Baldauf2013, Ginzburg2017, Liu2021, Friedrich2022}. In N-body simulations where the masses of dark matter haloes are known, \cite{Hamaus2010} and \cite{Cai2011} have shown that a linear combination of haloes of different masses with different weights can significantly reduce stochasticity. The primary information used for the improvement is the mass-dependence of halo power spectra $P_h(k,M)$, where $M$ is the mass of haloes. One way to understand this is that the linear halo bias $b(M)$ increases with $M$; for the same number of haloes, the signal-to-noise of $P_h(k, M)$ is higher for haloes with larger masses. Therefore, it is preferable to up-weight the high-mass haloes to increase the signal-to-noise. This is indeed the case for the optimal weights found in \citep{Hamaus2010, Cai2011}.  

However, this can not be the full physical picture. In addition to $M$, the clustering of haloes also depends on other secondary properties. Among them, the formation time of haloes is known to affect the halo clustering. For low-mass haloes, those formed earlier tend to exhibit stronger clustering than younger ones. This is known as the halo assembly bias \citep{Gao2005,Gao2007}. This indicates that there is room for further improvement in the reconstruction if this secondary property of haloes can be used, and it is the focus of this paper. 

A main challenge for this is that the formation time of haloes or galaxies is difficult to measure in observations. However, it is known to correlate with other properties of haloes, especially with halo concentration, tidal force, and their cosmic-web environments \citep[e.g.][]{Hahn2007, Skibba2011, Zhao2015, Chen2020, Balaguera-Antolinez2023}, with some arguing that cosmic-web anisotropy is the main indicator for halo assembly bias \citep{Borzyszkowski2017, Ramakrishnan2019, Mansfield2020}, the latter is accessible in observations. We will, therefore, focus on using the cosmic-web environment for the reconstruction.

Recent work of \citet{Balaguera-Antolinez2020} and \citet{Kitaura2022} has provided a model for reconstructing the matter density field by including local, non-local and higher-order bias terms, and showed their connections to the cosmic-web. They have demonstrated the success of the model using two-point and three-point statistics down to redshift $z=1$. In this study, we focus solely on two-point statistics and uses simple linear combinations of tracers i.e., no explicit inclusion of higher-order or non-local bias terms. In principle, our results will only apply to the large scales. As we will demonstrate, the method works as accurate for the reconstruction of two-point statistics at low redshifts. 

In Section (\ref{sec:method}) we outline our reconstruction method. In Section (\ref{sec:result}) we present our findings using N-body simulation. In the last section, we summarise our conclusions and discuss caveats and limitations of our method.
%%%%%%%%%%%%%%%%%%%{--
\section{Method}\label{sec:method}
Haloes are discrete and biased tracers of the underlying continuous matter distribution, described by $\rho_m = \bar{\rho}_m(1+\delta_m)$, where $\bar{\rho}_m$ denotes the mean matter density, we can define the halo density field using the Dirac delta function $\delta_D$ as $\rho_h = \sum_{i}^{N} w_i \delta_D(\vec{x} - \vec{x}_i) = \bar{\rho}_h(1 + \delta_h)$. Here, $N$ represents the number of haloes within the simulation box with a volume $V_{\text{BOX}}$, $w_i$ denotes the weight of each halo, and $\bar{\rho}_h = \sum_{i}^{N} w_i/V_{\text{BOX}}$ is the mean halo density. Depending on the scenario, the weight of each halo can be set to $1$ for equal weighting or to the halo mass for mass weighting. We will distinguish these two scenarios using superscripts `$N$' and `$M$', such as $\delta_h^N$ and $\delta_h^M$.

Given a smoothing scale (assuming a Top-hat window unless specified), either scenario ($\delta_h^N$ or $\delta_h^M$) yields deterministic correlation coefficients with matter overdensity $\delta_m$. If we conceptualise the original halo catalog as a composite of $N_s$ sub-catalogs within the same simulation box, the representation of the halo density field can be formulated as $\rho_h = \sum_{i}^{N_s} \rho_i$, where $\rho_i = \sum_j^{N_i} w_j\delta_D(\vec{x}-\vec{x}_j)$ represents the $i$th sub-catalog, with $N_i$ denoting the number of haloes in it. Then, a linear combination of these $N_s$ sub-catalogs yields a reconstructed density field $\rho_c = \sum_i^{N_s} W_i \rho_i$, where $W_i \in \mathbb{R}$ is a free parameter that will be optimised. It can be demonstrated that the resulting overdensity field $\delta_c$ is expressed as: 
\begin{equation}
    \delta_c \equiv \frac{\rho_c}{\bar{\rho}_c} - 1 = \frac{\sum_i^{N_s} W_i\bar{\rho}_i(1+\delta_i)}{\sum_i^{N_s} W_i\bar{\rho}_i}-1 = \bm{\omega}^T\bm{\delta},
    \label{eq:rc}
\end{equation}
where $\bar{\rho}_i=\sum_j^{N_i} w_j/V_{\text{BOX}}$ is the mean density of the $i$th sub-catalog, $\bm\omega$ and $\bm\delta$ are $N_s$-vectors, with elements $\omega_i = \frac{W_i\bar{\rho}_i}{\sum W_i\bar{\rho}_i}$ and $\delta_i=\frac{\rho_i}{\bar{\rho}_i}-1$ respectively. 
%%%%%%%%%%%%%%%%%%%%--}

Combining equations \eqref{eq:cross} and \eqref{eq:rc}, we have:
\begin{equation}
    r_c^2 = \frac{\sum_{ij}\omega_iB_{ij}\omega_j}{\sum_{ij}\omega_iC_{ij}\omega_j} =\frac{\bm\omega^T\bm{B}\bm{\omega}}{\bm\omega^T\bm{C}\bm{\omega}},
\end{equation}
with the symmetric matrices $\bm{B}$ and $\bm{C}$ defined as
\begin{equation}
    \left\{
    \begin{aligned}
        B_{ij} &= \frac{\langle\delta_i\delta_m\rangle\langle\delta_j\delta_m\rangle}{\langle\delta_m\delta_m\rangle}, \\
        C_{ij} &= \langle\delta_i\delta_j\rangle.
    \end{aligned}
    \right.
\end{equation}
To solve for the weights that maximize $r_c^2$, we set the derivative of $r_c^2$ to be $0$ to obtain:
\begin{equation}
    \bm{B}\bm\omega = r_c^2\bm{C}\bm\omega.
    \label{eq:eigen2}
\end{equation}

\begin{figure*}
    \centering
    \includegraphics[width=1.0\textwidth]{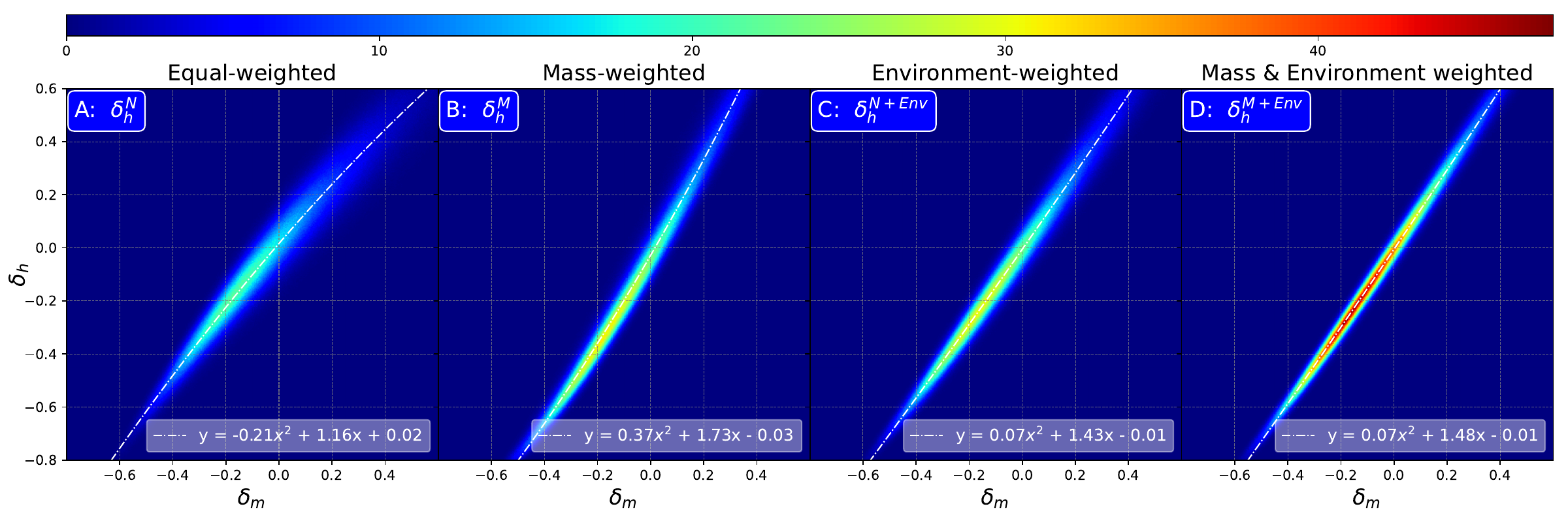}
    \caption{Joint probability density distributions of matter density fields $\delta_m$ and reconstructed halo density field with different weights. Each panel contains a random sampling of $\sim10^7$ points within the simulation box, smoothed by a top-hat filter with the radius of $R=30~ h^{-1}\text{Mpc}$. The coloursbar indicates the particle number density. Panel A: halo number density $\delta_h^N$ (equal weighting) versus $\delta_m$. B: Mass-weighted halo density $\delta_h^M$ versus $\delta_m$. C: Optimal-environmental weighted halo density $\delta_h^{N+Env}$ versus $\delta_m$. D: $\delta_h^{M+Env}$ versus $\delta_m$: optimally weighted halo density field with four halo mass bins and four environmental bins. The threshold parameter for the environmental split has been optimised, with $\lambda_{\text{th}} = 11.25$ and $17.5$ for C and D, respectively. See also Figure~\ref{fig:rc_L_th} and the main text for the details. Quadratic fits with the best-fit parameters shown in the legends.} 
    \label{fig:scatterPlot} 
\end{figure*}
The above eigenvalue equation (\ref{eq:eigen2}) is a generalized eigen-problem that can be solved numerically. We use the linear algebra library LAPACK\footnote{\url{https://netlib.org/lapack}}\href{https://netlib.org/lapack/explore-html/dc/dd2/group__double_o_t_h_e_reigen_ga4e4203d1260f4deffe7679ac49af4f10.html}{\text{::dspgv()}} to do this. The square root of the maximum eigenvalue corresponds to the maximum correlation coefficient, and the corresponding normalized eigenvector represents the optimal weight $\bm{\omega}$. The key information needed for solving the above equation is the matrices of $\bm{C}$ and $\bm{B}$. These can be measured from N-body simulations. 

In summary, our method for the reconstruction is to maximize the correlation coefficients between the halo field and the matter density field. To do that, we make a linear combination of the tracers; each is given a weight. We can solve for the weights $\bm{\omega}$ that maximize the correlation coefficients. The above derivation follows closely that of \cite{Cai2011} but using a different target function $r^2$. The results agree with each other. The derivation is general i.e., given any field of tracers, which can be dark matter haloes, galaxies, or HI field, and its underlying matter field $\delta_m$, we can solve equation (\ref{eq:eigen2}) to find the optimal $\bm{\omega}$ that maximizes $r_c$. Note again that by definition, our optimisation applies to two-point statistics only. When the field is non-Gaussian, there is no guarantee that the reconstruction is optimal for beyond-two-point statistics.
Next, we will apply this method to haloes in simulations, using information about their cosmic-web environment. 

\begin{figure}
    \centering
    \includegraphics[width=\columnwidth]{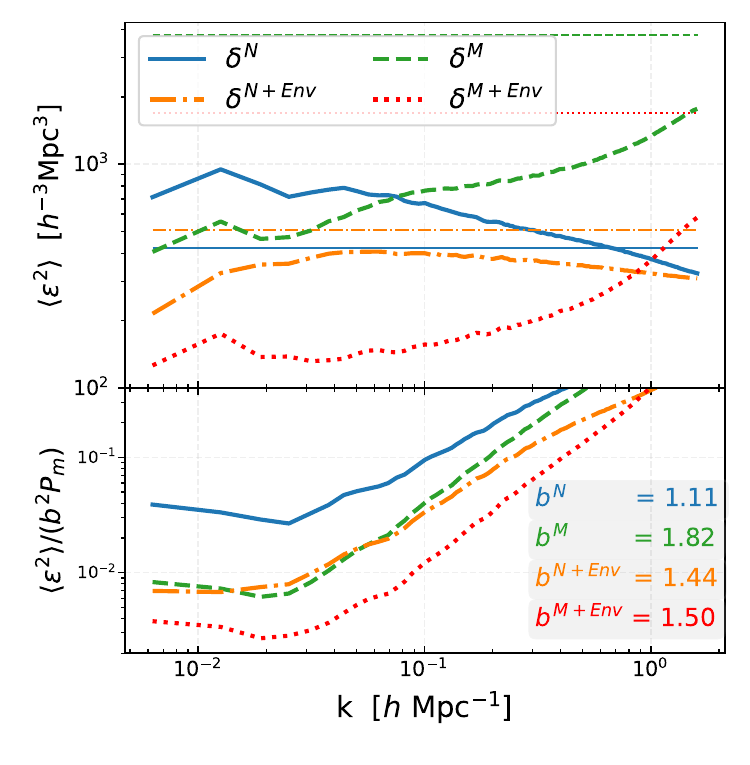}
    \caption{Upper panel: the power spectra of the noise $\langle \epsilon^2\rangle=\langle(b\delta_m-\delta_h)^2\rangle$ for the four reconstructed halo fields shown in Figure~\ref{fig:scatterPlot} and labeled in the legend. $\delta_h^N$:equal weighting; $\delta_h^M$: mass weighting; $\delta_h^N+Env$: environmental weighting; $\delta_h^M+Env$: optimal mass and environmental weighting.
    Their corresponding shot-noise levels are shown in horizontal lines of the same colours. 
    Bottom: the corresponding fractional errors on the power spectra, or the `noise-to-signal' ratio $\langle \epsilon^2\rangle/(b^2P_m)$, see also equation~(\ref{eq:S/N}). } 
    \label{fig:noise_R30}
\end{figure}

\section{Results from N-body simulations}\label{sec:result}
 We use the public {\sc MultiDark} MDPL2 simulations \cite{Klypin2016} and their corresponding {\sc rockstar} halo catalogues \citep{Behroozi2013} for our analysis. The simulation is run with $3840^3$ particles in a box of $1\,h^{-1}$Gpc following a flat $\Lambda$CDM model with $\Omega_m=0.307,\Omega_b=0.048,h=0.678,n_s=0.96,\sigma_8=0.823$. We use a sample of $\sim2.37\times10^6$ haloes with the minimal halo mass of $2.0\times 10^{12}\, h^{-1}\rm{M}_{\odot}$. 
 
 We first split the haloes sample into equal-number bins according to their masses. We then further split each mass bin according to their cosmic-web environment-- voids, sheets, filaments, and knots. We define the environment using the Hessian of the gravitational potential \citep{Hahn2007,Forero-Romero2009} smoothed at $1\,h^{-1}$Mpc. The three eigenvalues of the field ($\lambda_{1,2,3}$) are compared with a threshold value $\lambda_\text{th}$ to classify each volume as voids ($\lambda_{1,2,3}<\lambda_\text{th}$), sheets ($\lambda_{1,2}<\lambda_\text{th}< \lambda_{3}$), filaments ($\lambda_{1}<\lambda_\text{th}< \lambda_{2,3}$), and knots ($\lambda_\text{th}< \lambda_{1,2,3}$) \citep{Forero-Romero2009}. The threshold value of $\lambda_\text{th}$ is a free parameter that will be commented later. With this, the halo sample is split into four environmental bins according to their host cosmic-web environment. So, the halo field has two dependencies: mass $M$ and environment $Env$. The auto- and cross-correlations of $\delta_h(M, Env)$ form the matrix of $\bm{C}$ needed in equation~(\ref{eq:eigen2}). We have four bins of the environment by definition. If we choose four mass bins, $\bm{C}$ will be a matrix of $16\times16$. Its correlation with the matter density field $\delta_m$ yields the matrix $\bm{B}$. These will be inserted into equation~(\ref{eq:eigen2}) to find the maximum $r_c$ and the optimal weights ${\bm \omega}$.

Figure~\ref{fig:scatterPlot} presents comparisons of scatter plots between the dark matter density field versus the reconstructed halo field with different weights. We can see that the scatter is the largest when haloes are weighted equally (panel A), followed by mass-weighting (panel B), and then the optimisation with cosmic-web information (panel C). It is clear that by optimising with the cosmic-web information, we can reduce the stochasticity to a level that is comparable with the mass weighting case. Panel D shows the result from optimising with $M$ and $Env$, which reduces the stochasticity further to achieve the lowest level of all. Therefore, the information from $M$ and $Env$ are complementary. Their combination is better than having each of them alone. 

It is also worth noting that the mean correlation relation between the halo field and the mass field, which is the deterministic component of the bias, also tends to be more linear for the optimal weighting cases, i.e., having the smallest quadratic coefficient $b_2$ when fitted with a second-order polynomial function ($b_2=-0.21, 0.37, 0.07, 0.07$ for the four cases shown in the legend of the figure). This suggests that the second-order bias of haloes is suppressed by the optimisation -- an unexpected but preferable outcome. This is consistent with the results report in \citep{Cai2011}. 

The reduction of scatter in Figure~\ref{fig:scatterPlot} can be translated into an increase of signal-to-noise for halo clustering. The upper panel of Figure~\ref{fig:noise_R30} compares the noise spectra for the above four cases. Note that the Shot-noise levels are different for different weighting schemes. We can see that except equal-weighting where $\langle\epsilon^2\rangle$ is comparable to shot-noise, all the other noise spectra are below their shot-noise levels. By weighting with the cosmic-web environment (orange line), we are able to bring the noise level down below the case of mass weighting (green line). When we incorporate both $M$ and $Env$ for the optimisation, the noise level is the lowest, approximately one order of magnitude below the shot-noise level. This is $\sim$75\% lower than using the information of $M$ alone.

The lower panel of Figure~\ref{fig:noise_R30} compares $\langle\epsilon^2\rangle/(b^2P_m)$ -- the fractional errors on the power spectra, or the `noise-to-signal', for all the four cases. This is directly related to $r_c$ through equation~(\ref{eq:S/N}). We can see again that environmental weighting alone is comparable to mass weighting. They are approximately three times lower than equal weighting at low-$k$'s where the baryon acoustic oscillations are. optimising with both $M$ and $Env$ reduces the ratio further by another factor of $\sim$2-3. The optimal weighting is thus expected to increase the signal-to-noise ratio of halo clustering by the same amount in the regime where shot-noise dominates.

\begin{figure}
    \centering
    \includegraphics[width=\columnwidth]{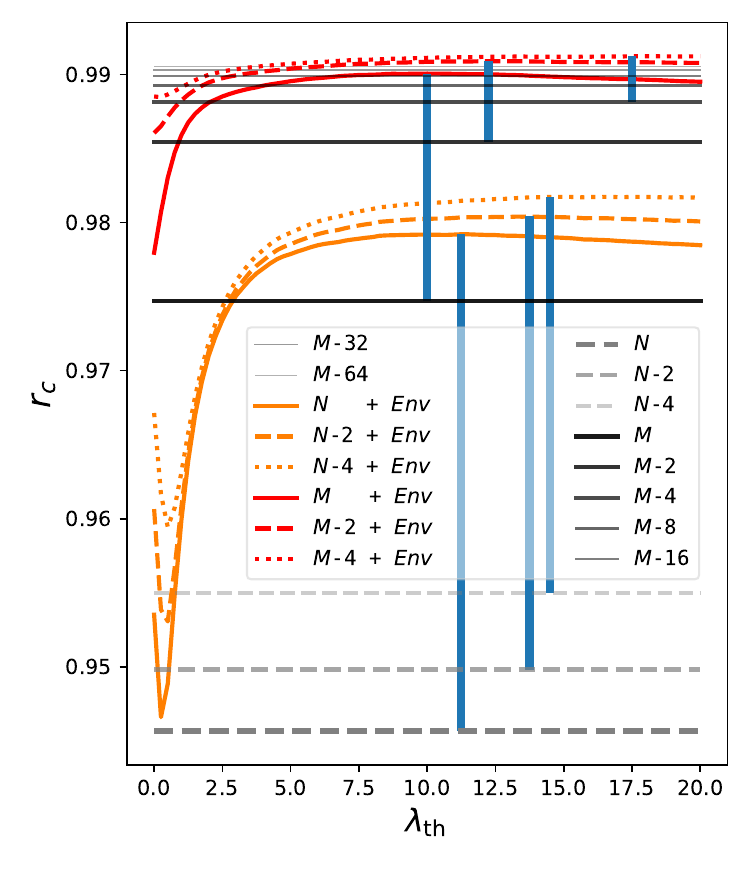}
    \caption{The correlation coefficient $r_c$ between the matter density and reconstructed fields versus the threshold value $\lambda_{\text{th}}$ for cosmic-web classification. Orange lines ($N$): equal-weighting plus environmental weighting; red lines ($M$): mass-weighting plus environmental weighting. The horizontal lines of dashed gray and solid black represent results from mass weighting and equal weighting only, respectively, without any environmental information. Blue vertical lines indicate the peak $r_c$ values and their length indicates the impact of the environment. The measurements are performed in configuration space with a $R=15\, h^{-1}\text{Mpc}$ top-hat smoothing.} 
    \label{fig:rc_L_th}
\end{figure}
The level of improvement for the reconstruction by having the information about the environment depends on the number of halo mass bins ${M}_\text{bin}$ and the threshold value $\lambda_\text{th}$ for the cosmic-web classification. We discuss these two variables as follows.

For a fixed ${M}_\text{bin}$, $r_c$ varies  with $\lambda_\text{th}$, as shown in orange and red lines in Figure~\ref{fig:rc_L_th}. There is an optimal $\lambda_\text{th}$ at which $r_c$ is maximized. We find that the optimal $\lambda_\text{th}$ is typically larger than zero. This makes both the volume fraction for voids and the number of haloes classified as void-haloes the largest, followed by sheets, filaments, and knots 
(see also figures in Supplementary Material and \citep{Hahn2007, Forero-Romero2009}). 
We have optimised $\lambda_\text{th}$ for results in the paper unless specified. We present the best-fit functions for the optimal $\lambda_\text{th}$ versus ${M}_\text{bin}$ and the corresponding $r_c$ versus ${M}_\text{bin}$ in Supplementary Material. 

With the optimal $\lambda_\text{th}$, $r_c$ increases with increasing ${M}_\text{bin}$ (horizontal lines in Figure~\ref{fig:rc_L_th}). When ${M}_\text{bin}$ is large, each individual halo is effectively in a single bin. All properties of haloes are used in the optimisation, leaving no room for further improvement. In the other extreme case where no information about the halo mass is known, i.e., ${M}_\text{bin}=1$, the improvement for the reconstruction with the cosmic-web information is maximized. This is illustrated in Figure~\ref{fig:rc_L_th} by the length of the blue vertical line in terms of the correlation coefficient $r_c$. We can see that equal-weighting provides the lowest $r_c$ (gray dashed line, labeled as $N$); using the cosmic-web information alone (orange line, labeled as $N+Env$) significantly improves $r_c$, and slightly outperforms the mass weighting case (labeled as $M$); having two or more mass bin for the optimisation increases $r_c$ further. $r_c=0.99$ is achieved by having one mass bin and the environment information ($M+Env$, red-solid line). This is approximately the same as having 16 mass bins ($M$-16) for optimisation. Note that having many mass bins is unrealistic, given the challenge of measuring the mass of haloes in observations. The gain we can achieve using the cosmic-web information is, therefore, highly complementary. 

It is worth noting that when optimising with the mass information, we assume that the mass of each individual halo is known. This is unrealistic in practice. We anticipate large uncertainties in estimating halo mass in real observations. Therefore, the benefit of using environmental weighting may be more prominent than we have shown. 

Since the Hessian matrix of the potential, which contains information about the density field, was used for the cosmic-web classification, there may be concerns that the information about the density field was over used for the reconstruction. Note that we only use the tidal field to classify haloes into four different kinds: voids, sheets, filaments and knots i.e., to draw boundaries. These provides four variables -- weights. They are let free to vary to minimise our loss function. So the actual values of eigenvalues of the Hessian matrix have never been used in the optimisation.

\section{Conclusions and discussion}\label{sec:conclusion}
In summary, we have found significant improvement in the correlation between haloes and their underlying mass field using information about the cosmic-web environment. The level of improvement for the reconstruction is at least as good as optimising using halo mass. Environment and halo mass complement each other. Their combination yields the best reconstruction than having each of them individually. 

These results are in qualitative agreement with what was
presented in \citep{Balaguera-Antolinez2020, Kitaura2022}, both showing the importance of having information about the cosmic web for the reconstruction. In terms of the cross-correlation coefficients between the reconstructed halo field and the matter density field, both studies reach the order of 10\% at $k\sim 1$Mpc/$h$. Exact comparison is difficult due to the difference for the minimal halo mass of the two catalogues and the redshift at which the measurements were conducted. 
In addition, an analytical formula including higher order and non-local biases were provided, the connections between their bias terms and the cosmic web were shown explicitly in \citep{Kitaura2022}. 

These findings have direct observational implications.

Reducing stochasticity, or noise, for the galaxy/halo population will directly increase the signal-to-noise for galaxy clustering. This is crucial for precise measurement of the BAO, especially at high redshifts where the tracers are sparse \citep[e.g.][]{Seo2007, White2010,Cohn2016, Lepori2017,Ding2018, Seo2022}. Galaxy redshift surveys such as BOSS \citep{BOSS2013}, eBOSS \citep{Alam2021} and DESI \citep{DESI2016}, are designed to target at a specific type of galaxy, such as LRG, galaxies with approximately constant stellar mass, and emission line galaxies, but it is challenging to know the mass of individual haloes. This limits the potential of shot-noise reduction using the halo mass information. In contrast, the cosmic-web information is more readily available from computing the Hessian matrix using the galaxy samples themselves.

Reducing stochasticity between galaxies/haloes versus the matter field also has direct benefits for forward modeling, e.g. {\sc Borg} \citep{Jasche2019}, constrained simulations \citep[e.g.][]{Wang2014}, and cosmological inference at the field level using observations of the large-scale structure \citep[e.g.][]{Schmittfull2019,Zhou2023}. The common starting point for the above analyses is the observed galaxy number density field $\delta_g$; one then tries to find the initial matter density field $\delta_m^i$ which, after non-linear evolution, may generate the observed $\delta_g$. Reducing stochasticity for $\delta_g$ will directly reduce the errors between the observational constraints with their initial conditions. Our method opens a promising path for achieving this.

To this end, our study remains theoretical and idealised. The key information needed for the reconstruction, the matrix of ${\bf B}$ and ${\bf C}$ is assumed to be known perfectly from simulations in this study. 
Also, the classification of the cosmic-web starts from computing the gravitational potential, which we uses dark matter particles for it. It remains to be explore to what extent the gain of information from reconstruction will be degraded in real observations when imperfect information and possible observational systematics come to play.   

In real observations, the smoothing scale will be limited by the number density of galaxies, although we have tested using a relatively large smoothing scale ($R=5\, h^{-1}\text{Mpc}$) to define the cosmic web, and there is still an obvious gain of information. 

An additional challenge is that the observed galaxies are in redshift space, so the defined cosmic web may differ from its real-space version. Further investigations are needed to fully realise the method's strength in observations. Despite these challenges, recent observational analyses have suggested possible impact of environmental effect on galaxies and haloes \citep[e.g.][]{Calderon2018,Alam2020}

We also caution that our study focuses solely on two-point statistics, it remains to be explored if our simple linear combination of weighted halo field can also accurately capture higher order statistics of the matter field. This has indeed been demonstrated to be the case using the bias model in \citep{Kitaura2022}.

We are aware that the environment is not the only secondary property of haloes that can be useful for reconstruction. We have explored all other halo properties typically defined in N-body simulations, finding that information from the tidal force and halo concentration can also help to improve the reconstruction. This is expected as these halo properties are correlated and are all related to the assembly bias of haloes  \citep[e.g.][]{Hahn2007, Chen2020}. 

%{\bf We note that the very recent work by \cite{Balaguera-Antolinez2023} also incorporates environmental information with other intrinsic halo properties to investigate the clustering of dark matter haloes. Their study arrives at similar conclusions, highlighting the significance of the tidal field as a secondary property in clustering analysis.}

\section*{Acknowledgements}
The CosmoSim database used in this paper is a service by the Leibniz-Institute for Astrophysics Potsdam (AIP). The MultiDark database was developed in cooperation with the Spanish MultiDark Consolider Project CSD2009-00064.
The authors gratefully acknowledge the Gauss Centre for Supercomputing e.V. (\url{www.gauss-centre.eu}) and the Partnership for Advanced Supercomputing in Europe (PRACE, \url{www.prace-ri.eu}) for funding the MultiDark simulation project by providing computing time on the GCS Supercomputer SuperMUC at Leibniz Supercomputing Centre (LRZ, \url{www.lrz.de}). FLL is supported by the National Key R\&D Program of China through grant 2020YFC2201400 and the Key Program of NFSC through grant 11733010 and 11333008. YC acknowledges the support of
the Royal Society through a University Research Fellowship. For the purpose of open access, the author has applied a Creative Commons Attribution (CC BY) licence to any Author Accepted Manuscript version arising from this submission.
We thank the anonymous referee for their very useful comments, which have helped to improve the clarity of the paper.

\section*{Data Availability}
All the simulation data used in this paper is available through the MultiDark database website \url{https://www.cosmosim.org}. 
The results produced in this work are available upon reasonable request to the authors.

\bibliographystyle{mnras}
\bibliography{main.bib}

\begin{thebibliography}{}
\makeatletter
\relax
\def\mn@urlcharsother{\let\do\@makeother \do\$\do\&\do\#\do\^\do\_\do\%\do\~}
\def\mn@doi{\begingroup\mn@urlcharsother \@ifnextchar [ {\mn@doi@} {\mn@doi@[]}}
\def\mn@doi@[#1]#2{\def\@tempa{#1}\ifx\@tempa\@empty \href {http://dx.doi.org/#2} {doi:#2}\else \href {http://dx.doi.org/#2} {#1}\fi \endgroup}
\def\mn@eprint#1#2{\mn@eprint@#1:#2::\@nil}
\def\mn@eprint@arXiv#1{\href {http://arxiv.org/abs/#1} {{\tt arXiv:#1}}}
\def\mn@eprint@dblp#1{\href {http://dblp.uni-trier.de/rec/bibtex/#1.xml} {dblp:#1}}
\def\mn@eprint@#1:#2:#3:#4\@nil{\def\@tempa {#1}\def\@tempb {#2}\def\@tempc {#3}\ifx \@tempc \@empty \let \@tempc \@tempb \let \@tempb \@tempa \fi \ifx \@tempb \@empty \def\@tempb {arXiv}\fi \@ifundefined {mn@eprint@\@tempb}{\@tempb:\@tempc}{\expandafter \expandafter \csname mn@eprint@\@tempb\endcsname \expandafter{\@tempc}}}

\bibitem[\protect\citeauthoryear{{Alam}, {Peacock}, {Kraljic}, {Ross}  \& {Comparat}}{{Alam} et~al.}{2020}]{Alam2020}
{Alam} S.,  {Peacock} J.~A.,  {Kraljic} K.,  {Ross} A.~J.,   {Comparat} J.,  2020, \mn@doi [\mnras] {10.1093/mnras/staa1956}, \href {https://ui.adsabs.harvard.edu/abs/2020MNRAS.497..581A} {497, 581}

\bibitem[\protect\citeauthoryear{{Alam} et~al.}{{Alam} et~al.}{2021}]{Alam2021}
{Alam} S.,  et~al., 2021, \mn@doi [\prd] {10.1103/PhysRevD.103.083533}, \href {https://ui.adsabs.harvard.edu/abs/2021PhRvD.103h3533A} {103, 083533}

\bibitem[\protect\citeauthoryear{{Angulo}, {Baugh}, {Frenk}  \& {Lacey}}{{Angulo} et~al.}{2008}]{Angulo2008}
{Angulo} R.~E.,  {Baugh} C.~M.,  {Frenk} C.~S.,   {Lacey} C.~G.,  2008, \mn@doi [\mnras] {10.1111/j.1365-2966.2007.12587.x}, \href {https://ui.adsabs.harvard.edu/abs/2008MNRAS.383..755A} {383, 755}

\bibitem[\protect\citeauthoryear{{Balaguera-Antol{\'\i}nez} et~al.,}{{Balaguera-Antol{\'\i}nez} et~al.}{2020}]{Balaguera-Antolinez2020}
{Balaguera-Antol{\'\i}nez} A.,  et~al., 2020, \mn@doi [\mnras] {10.1093/mnras/stz3206}, \href {https://ui.adsabs.harvard.edu/abs/2020MNRAS.491.2565B} {491, 2565}

\bibitem[\protect\citeauthoryear{{Balaguera-Antol{\'\i}nez}, {Montero-Dorta}  \& {Favole}}{{Balaguera-Antol{\'\i}nez} et~al.}{2023}]{Balaguera-Antolinez2023}
{Balaguera-Antol{\'\i}nez} A.,  {Montero-Dorta} A.~D.,   {Favole} G.,  2023, \mn@doi [arXiv e-prints] {10.48550/arXiv.2311.12991}, \href {https://ui.adsabs.harvard.edu/abs/2023arXiv231112991B} {p. arXiv:2311.12991}

\bibitem[\protect\citeauthoryear{{Baldauf}, {Seljak}, {Smith}, {Hamaus}  \& {Desjacques}}{{Baldauf} et~al.}{2013}]{Baldauf2013}
{Baldauf} T.,  {Seljak} U.,  {Smith} R.~E.,  {Hamaus} N.,   {Desjacques} V.,  2013, \mn@doi [\prd] {10.1103/PhysRevD.88.083507}, \href {https://ui.adsabs.harvard.edu/abs/2013PhRvD..88h3507B} {88, 083507}

\bibitem[\protect\citeauthoryear{{Behroozi}, {Wechsler}  \& {Wu}}{{Behroozi} et~al.}{2013}]{Behroozi2013}
{Behroozi} P.~S.,  {Wechsler} R.~H.,   {Wu} H.-Y.,  2013, \mn@doi [\apj] {10.1088/0004-637X/762/2/109}, \href {https://ui.adsabs.harvard.edu/abs/2013ApJ...762..109B} {762, 109}

\bibitem[\protect\citeauthoryear{{Beutler} et~al.,}{{Beutler} et~al.}{2014}]{Beutler2014}
{Beutler} F.,  et~al., 2014, \mn@doi [\mnras] {10.1093/mnras/stu1051}, \href {https://ui.adsabs.harvard.edu/abs/2014MNRAS.443.1065B} {443, 1065}

\bibitem[\protect\citeauthoryear{{Bonoli} \& {Pen}}{{Bonoli} \& {Pen}}{2009}]{Bonoli2009}
{Bonoli} S.,  {Pen} U.~L.,  2009, \mn@doi [\mnras] {10.1111/j.1365-2966.2009.14829.x}, \href {https://ui.adsabs.harvard.edu/abs/2009MNRAS.396.1610B} {396, 1610}

\bibitem[\protect\citeauthoryear{{Borzyszkowski}, {Porciani}, {Romano-D{\'\i}az}  \& {Garaldi}}{{Borzyszkowski} et~al.}{2017}]{Borzyszkowski2017}
{Borzyszkowski} M.,  {Porciani} C.,  {Romano-D{\'\i}az} E.,   {Garaldi} E.,  2017, \mn@doi [\mnras] {10.1093/mnras/stx873}, \href {https://ui.adsabs.harvard.edu/abs/2017MNRAS.469..594B} {469, 594}

\bibitem[\protect\citeauthoryear{{Cai} \& {Bernstein}}{{Cai} \& {Bernstein}}{2012}]{Cai2012}
{Cai} Y.-C.,  {Bernstein} G.,  2012, \mn@doi [\mnras] {10.1111/j.1365-2966.2012.20676.x}, \href {https://ui.adsabs.harvard.edu/abs/2012MNRAS.422.1045C} {422, 1045}

\bibitem[\protect\citeauthoryear{{Cai}, {Bernstein}  \& {Sheth}}{{Cai} et~al.}{2011}]{Cai2011}
{Cai} Y.-C.,  {Bernstein} G.,   {Sheth} R.~K.,  2011, \mn@doi [\mnras] {10.1111/j.1365-2966.2010.17969.x}, \href {https://ui.adsabs.harvard.edu/abs/2011MNRAS.412..995C} {412, 995}

\bibitem[\protect\citeauthoryear{{Calderon}, {Berlind}  \& {Sinha}}{{Calderon} et~al.}{2018}]{Calderon2018}
{Calderon} V.~F.,  {Berlind} A.~A.,   {Sinha} M.,  2018, \mn@doi [\mnras] {10.1093/mnras/sty2000}, \href {https://ui.adsabs.harvard.edu/abs/2018MNRAS.480.2031C} {480, 2031}

\bibitem[\protect\citeauthoryear{{Chan} \& {Scoccimarro}}{{Chan} \& {Scoccimarro}}{2012}]{Chan2012}
{Chan} K.~C.,  {Scoccimarro} R.,  2012, \mn@doi [\prd] {10.1103/PhysRevD.86.103519}, \href {https://ui.adsabs.harvard.edu/abs/2012PhRvD..86j3519C} {86, 103519}

\bibitem[\protect\citeauthoryear{{Chen}, {Mo}, {Li}, {Wang}, {Yang}, {Zhang}  \& {Wang}}{{Chen} et~al.}{2020}]{Chen2020}
{Chen} Y.,  {Mo} H.~J.,  {Li} C.,  {Wang} H.,  {Yang} X.,  {Zhang} Y.,   {Wang} K.,  2020, \mn@doi [\apj] {10.3847/1538-4357/aba597}, \href {https://ui.adsabs.harvard.edu/abs/2020ApJ...899...81C} {899, 81}

\bibitem[\protect\citeauthoryear{{Cohn}, {White}, {Chang}, {Holder}, {Padmanabhan}  \& {Dor{\'e}}}{{Cohn} et~al.}{2016}]{Cohn2016}
{Cohn} J.~D.,  {White} M.,  {Chang} T.-C.,  {Holder} G.,  {Padmanabhan} N.,   {Dor{\'e}} O.,  2016, \mn@doi [\mnras] {10.1093/mnras/stw108}, \href {https://ui.adsabs.harvard.edu/abs/2016MNRAS.457.2068C} {457, 2068}

\bibitem[\protect\citeauthoryear{{DESI Collaboration} et~al.}{{DESI Collaboration} et~al.}{2016}]{DESI2016}
{DESI Collaboration} et~al., 2016, \mn@doi [arXiv e-prints] {10.48550/arXiv.1611.00036}, \href {https://ui.adsabs.harvard.edu/abs/2016arXiv161100036D} {p. arXiv:1611.00036}

\bibitem[\protect\citeauthoryear{{Dawson} et~al.}{{Dawson} et~al.}{2013}]{BOSS2013}
{Dawson} K.~S.,  et~al., 2013, \mn@doi [\aj] {10.1088/0004-6256/145/1/10}, \href {https://ui.adsabs.harvard.edu/abs/2013AJ....145...10D} {145, 10}

\bibitem[\protect\citeauthoryear{{Dekel} \& {Lahav}}{{Dekel} \& {Lahav}}{1999}]{Dekel1999}
{Dekel} A.,  {Lahav} O.,  1999, \mn@doi [\apj] {10.1086/307428}, \href {https://ui.adsabs.harvard.edu/abs/1999ApJ...520...24D} {520, 24}

\bibitem[\protect\citeauthoryear{{Ding}, {Seo}, {Vlah}, {Feng}, {Schmittfull}  \& {Beutler}}{{Ding} et~al.}{2018}]{Ding2018}
{Ding} Z.,  {Seo} H.-J.,  {Vlah} Z.,  {Feng} Y.,  {Schmittfull} M.,   {Beutler} F.,  2018, \mn@doi [\mnras] {10.1093/mnras/sty1413}, \href {https://ui.adsabs.harvard.edu/abs/2018MNRAS.479.1021D} {479, 1021}

\bibitem[\protect\citeauthoryear{{Eriksen} \& {Gazta{\~n}aga}}{{Eriksen} \& {Gazta{\~n}aga}}{2018}]{Eriksen2018}
{Eriksen} M.,  {Gazta{\~n}aga} E.,  2018, \mn@doi [\mnras] {10.1093/mnras/sty2168}, \href {https://ui.adsabs.harvard.edu/abs/2018MNRAS.480.5226E} {480, 5226}

\bibitem[\protect\citeauthoryear{{Forero-Romero}, {Hoffman}, {Gottl{\"o}ber}, {Klypin}  \& {Yepes}}{{Forero-Romero} et~al.}{2009}]{Forero-Romero2009}
{Forero-Romero} J.~E.,  {Hoffman} Y.,  {Gottl{\"o}ber} S.,  {Klypin} A.,   {Yepes} G.,  2009, \mn@doi [\mnras] {10.1111/j.1365-2966.2009.14885.x}, \href {https://ui.adsabs.harvard.edu/abs/2009MNRAS.396.1815F} {396, 1815}

\bibitem[\protect\citeauthoryear{{Friedrich}, {Halder}, {Boyle}, {Uhlemann}, {Britt}, {Codis}, {Gruen}  \& {Hahn}}{{Friedrich} et~al.}{2022}]{Friedrich2022}
{Friedrich} O.,  {Halder} A.,  {Boyle} A.,  {Uhlemann} C.,  {Britt} D.,  {Codis} S.,  {Gruen} D.,   {Hahn} C.,  2022, \mn@doi [\mnras] {10.1093/mnras/stab3703}, \href {https://ui.adsabs.harvard.edu/abs/2022MNRAS.510.5069F} {510, 5069}

\bibitem[\protect\citeauthoryear{{Gao} \& {White}}{{Gao} \& {White}}{2007}]{Gao2007}
{Gao} L.,  {White} S. D.~M.,  2007, \mn@doi [\mnras] {10.1111/j.1745-3933.2007.00292.x}, \href {https://ui.adsabs.harvard.edu/abs/2007MNRAS.377L...5G} {377, L5}

\bibitem[\protect\citeauthoryear{{Gao}, {Springel}  \& {White}}{{Gao} et~al.}{2005}]{Gao2005}
{Gao} L.,  {Springel} V.,   {White} S. D.~M.,  2005, \mn@doi [\mnras] {10.1111/j.1745-3933.2005.00084.x}, \href {https://ui.adsabs.harvard.edu/abs/2005MNRAS.363L..66G} {363, L66}

\bibitem[\protect\citeauthoryear{{Ginzburg}, {Desjacques}  \& {Chan}}{{Ginzburg} et~al.}{2017}]{Ginzburg2017}
{Ginzburg} D.,  {Desjacques} V.,   {Chan} K.~C.,  2017, \mn@doi [\prd] {10.1103/PhysRevD.96.083528}, \href {https://ui.adsabs.harvard.edu/abs/2017PhRvD..96h3528G} {96, 083528}

\bibitem[\protect\citeauthoryear{{Hahn}, {Porciani}, {Carollo}  \& {Dekel}}{{Hahn} et~al.}{2007}]{Hahn2007}
{Hahn} O.,  {Porciani} C.,  {Carollo} C.~M.,   {Dekel} A.,  2007, \mn@doi [\mnras] {10.1111/j.1365-2966.2006.11318.x}, \href {https://ui.adsabs.harvard.edu/abs/2007MNRAS.375..489H} {375, 489}

\bibitem[\protect\citeauthoryear{{Hamaus}, {Seljak}, {Desjacques}, {Smith}  \& {Baldauf}}{{Hamaus} et~al.}{2010}]{Hamaus2010}
{Hamaus} N.,  {Seljak} U.,  {Desjacques} V.,  {Smith} R.~E.,   {Baldauf} T.,  2010, \mn@doi [\prd] {10.1103/PhysRevD.82.043515}, \href {https://ui.adsabs.harvard.edu/abs/2010PhRvD..82d3515H} {82, 043515}

\bibitem[\protect\citeauthoryear{{Hamaus}, {Seljak}  \& {Desjacques}}{{Hamaus} et~al.}{2012}]{Hamaus2012}
{Hamaus} N.,  {Seljak} U.,   {Desjacques} V.,  2012, \mn@doi [\prd] {10.1103/PhysRevD.86.103513}, \href {https://ui.adsabs.harvard.edu/abs/2012PhRvD..86j3513H} {86, 103513}

\bibitem[\protect\citeauthoryear{{Jasche} \& {Lavaux}}{{Jasche} \& {Lavaux}}{2019}]{Jasche2019}
{Jasche} J.,  {Lavaux} G.,  2019, \mn@doi [\aap] {10.1051/0004-6361/201833710}, \href {https://ui.adsabs.harvard.edu/abs/2019A&A...625A..64J} {625, A64}

\bibitem[\protect\citeauthoryear{{Junzhe Zhou} \& {Dodelson}}{{Junzhe Zhou} \& {Dodelson}}{2023}]{Zhou2023}
{Junzhe Zhou} A.,  {Dodelson} S.,  2023, \mn@doi [arXiv e-prints] {10.48550/arXiv.2304.01387}, \href {https://ui.adsabs.harvard.edu/abs/2023arXiv230401387J} {p. arXiv:2304.01387}

\bibitem[\protect\citeauthoryear{{Kitaura}, {Yepes}  \& {Prada}}{{Kitaura} et~al.}{2014}]{Kitaura2014}
{Kitaura} F.~S.,  {Yepes} G.,   {Prada} F.,  2014, \mn@doi [\mnras] {10.1093/mnrasl/slt172}, \href {https://ui.adsabs.harvard.edu/abs/2014MNRAS.439L..21K} {439, L21}

\bibitem[\protect\citeauthoryear{{Kitaura} et~al.,}{{Kitaura} et~al.}{2016}]{Kitaura2016}
{Kitaura} F.~S.,  et~al., 2016, \mn@doi [\mnras] {10.1093/mnras/stv2826}, \href {https://ui.adsabs.harvard.edu/abs/2016MNRAS.456.4156K} {456, 4156}

\bibitem[\protect\citeauthoryear{{Kitaura}, {Balaguera-Antol{\'\i}nez}, {Sinigaglia}  \& {Pellejero-Ib{\'a}{\~n}ez}}{{Kitaura} et~al.}{2022}]{Kitaura2022}
{Kitaura} F.~S.,  {Balaguera-Antol{\'\i}nez} A.,  {Sinigaglia} F.,   {Pellejero-Ib{\'a}{\~n}ez} M.,  2022, \mn@doi [\mnras] {10.1093/mnras/stac671}, \href {https://ui.adsabs.harvard.edu/abs/2022MNRAS.512.2245K} {512, 2245}

\bibitem[\protect\citeauthoryear{{Klypin}, {Yepes}, {Gottl{\"o}ber}, {Prada}  \& {He{\ss}}}{{Klypin} et~al.}{2016}]{Klypin2016}
{Klypin} A.,  {Yepes} G.,  {Gottl{\"o}ber} S.,  {Prada} F.,   {He{\ss}} S.,  2016, \mn@doi [\mnras] {10.1093/mnras/stw248}, \href {https://ui.adsabs.harvard.edu/abs/2016MNRAS.457.4340K} {457, 4340}

\bibitem[\protect\citeauthoryear{{Lavaux} \& {Jasche}}{{Lavaux} \& {Jasche}}{2021}]{Lavaux&Jasche2021}
{Lavaux} G.,  {Jasche} J.,  2021, \mn@doi [arXiv e-prints] {10.48550/arXiv.2104.12992}, \href {https://ui.adsabs.harvard.edu/abs/2021arXiv210412992L} {p. arXiv:2104.12992}

\bibitem[\protect\citeauthoryear{{Lee} et~al.,}{{Lee} et~al.}{2022}]{Lee2022}
{Lee} S.,  et~al., 2022, \mn@doi [\mnras] {10.1093/mnras/stab3028}, \href {https://ui.adsabs.harvard.edu/abs/2022MNRAS.509.2033L} {509, 2033}

\bibitem[\protect\citeauthoryear{{Lepori}, {Di Dio}, {Viel}, {Baccigalupi}  \& {Durrer}}{{Lepori} et~al.}{2017}]{Lepori2017}
{Lepori} F.,  {Di Dio} E.,  {Viel} M.,  {Baccigalupi} C.,   {Durrer} R.,  2017, \mn@doi [\jcap] {10.1088/1475-7516/2017/02/020}, \href {https://ui.adsabs.harvard.edu/abs/2017JCAP...02..020L} {2017, 020}

\bibitem[\protect\citeauthoryear{{Liu}, {Yu}  \& {Li}}{{Liu} et~al.}{2021}]{Liu2021}
{Liu} Y.,  {Yu} Y.,   {Li} B.,  2021, \mn@doi [\apjs] {10.3847/1538-4365/abe868}, \href {https://ui.adsabs.harvard.edu/abs/2021ApJS..254....4L} {254, 4}

\bibitem[\protect\citeauthoryear{{Ma}, {Zhang}, {Yu}  \& {Qin}}{{Ma} et~al.}{2023}]{Ma2023}
{Ma} R.,  {Zhang} P.,  {Yu} Y.,   {Qin} J.,  2023, \mn@doi [arXiv e-prints] {10.48550/arXiv.2306.15177}, \href {https://ui.adsabs.harvard.edu/abs/2023arXiv230615177M} {p. arXiv:2306.15177}

\bibitem[\protect\citeauthoryear{{Mansfield} \& {Kravtsov}}{{Mansfield} \& {Kravtsov}}{2020}]{Mansfield2020}
{Mansfield} P.,  {Kravtsov} A.~V.,  2020, \mn@doi [\mnras] {10.1093/mnras/staa430}, \href {https://ui.adsabs.harvard.edu/abs/2020MNRAS.493.4763M} {493, 4763}

\bibitem[\protect\citeauthoryear{{Patej} \& {Eisenstein}}{{Patej} \& {Eisenstein}}{2018}]{Patej2018}
{Patej} A.,  {Eisenstein} D.~J.,  2018, \mn@doi [\mnras] {10.1093/mnras/sty870}, \href {https://ui.adsabs.harvard.edu/abs/2018MNRAS.477.5090P} {477, 5090}

\bibitem[\protect\citeauthoryear{{Ramakrishnan}, {Paranjape}, {Hahn}  \& {Sheth}}{{Ramakrishnan} et~al.}{2019}]{Ramakrishnan2019}
{Ramakrishnan} S.,  {Paranjape} A.,  {Hahn} O.,   {Sheth} R.~K.,  2019, \mn@doi [\mnras] {10.1093/mnras/stz2344}, \href {https://ui.adsabs.harvard.edu/abs/2019MNRAS.489.2977R} {489, 2977}

\bibitem[\protect\citeauthoryear{{Schmittfull}, {Simonovi{\'c}}, {Assassi}  \& {Zaldarriaga}}{{Schmittfull} et~al.}{2019}]{Schmittfull2019}
{Schmittfull} M.,  {Simonovi{\'c}} M.,  {Assassi} V.,   {Zaldarriaga} M.,  2019, \mn@doi [\prd] {10.1103/PhysRevD.100.043514}, \href {https://ui.adsabs.harvard.edu/abs/2019PhRvD.100d3514S} {100, 043514}

\bibitem[\protect\citeauthoryear{{Seljak} \& {Warren}}{{Seljak} \& {Warren}}{2004}]{Seljak2004}
{Seljak} U.,  {Warren} M.~S.,  2004, \mn@doi [\mnras] {10.1111/j.1365-2966.2004.08297.x}, \href {https://ui.adsabs.harvard.edu/abs/2004MNRAS.355..129S} {355, 129}

\bibitem[\protect\citeauthoryear{{Seljak}, {Hamaus}  \& {Desjacques}}{{Seljak} et~al.}{2009}]{Seljak2009}
{Seljak} U.,  {Hamaus} N.,   {Desjacques} V.,  2009, \mn@doi [\prl] {10.1103/PhysRevLett.103.091303}, \href {https://ui.adsabs.harvard.edu/abs/2009PhRvL.103i1303S} {103, 091303}

\bibitem[\protect\citeauthoryear{{Seo} \& {Eisenstein}}{{Seo} \& {Eisenstein}}{2007}]{Seo2007}
{Seo} H.-J.,  {Eisenstein} D.~J.,  2007, \mn@doi [\apj] {10.1086/519549}, \href {https://ui.adsabs.harvard.edu/abs/2007ApJ...665...14S} {665, 14}

\bibitem[\protect\citeauthoryear{{Seo}, {Ota}, {Schmittfull}, {Saito}  \& {Beutler}}{{Seo} et~al.}{2022}]{Seo2022}
{Seo} H.-J.,  {Ota} A.,  {Schmittfull} M.,  {Saito} S.,   {Beutler} F.,  2022, \mn@doi [\mnras] {10.1093/mnras/stac082}, \href {https://ui.adsabs.harvard.edu/abs/2022MNRAS.511.1557S} {511, 1557}

\bibitem[\protect\citeauthoryear{{Skibba} \& {Macci{\`o}}}{{Skibba} \& {Macci{\`o}}}{2011}]{Skibba2011}
{Skibba} R.~A.,  {Macci{\`o}} A.~V.,  2011, \mn@doi [\mnras] {10.1111/j.1365-2966.2011.19218.x}, \href {https://ui.adsabs.harvard.edu/abs/2011MNRAS.416.2388S} {416, 2388}

\bibitem[\protect\citeauthoryear{{Wang}, {Mo}, {Yang}, {Jing}  \& {Lin}}{{Wang} et~al.}{2014}]{Wang2014}
{Wang} H.,  {Mo} H.~J.,  {Yang} X.,  {Jing} Y.~P.,   {Lin} W.~P.,  2014, \mn@doi [\apj] {10.1088/0004-637X/794/1/94}, \href {https://ui.adsabs.harvard.edu/abs/2014ApJ...794...94W} {794, 94}

\bibitem[\protect\citeauthoryear{{White}}{{White}}{2010}]{White2010}
{White} M.,  2010, \mn@doi [arXiv e-prints] {10.48550/arXiv.1004.0250}, \href {https://ui.adsabs.harvard.edu/abs/2010arXiv1004.0250W} {p. arXiv:1004.0250}

\bibitem[\protect\citeauthoryear{{Yang}, {Zhang}, {Zhang}  \& {Yu}}{{Yang} et~al.}{2015}]{Yang2015}
{Yang} X.,  {Zhang} P.,  {Zhang} J.,   {Yu} Y.,  2015, \mn@doi [\mnras] {10.1093/mnras/stu2375}, \href {https://ui.adsabs.harvard.edu/abs/2015MNRAS.447..345Y} {447, 345}

\bibitem[\protect\citeauthoryear{{Zhao}, {Kitaura}, {Chuang}, {Prada}, {Yepes}  \& {Tao}}{{Zhao} et~al.}{2015}]{Zhao2015}
{Zhao} C.,  {Kitaura} F.-S.,  {Chuang} C.-H.,  {Prada} F.,  {Yepes} G.,   {Tao} C.,  2015, \mn@doi [\mnras] {10.1093/mnras/stv1262}, \href {https://ui.adsabs.harvard.edu/abs/2015MNRAS.451.4266Z} {451, 4266}

\bibitem[\protect\citeauthoryear{{Zhao} et~al.,}{{Zhao} et~al.}{2019}]{zhao2019}
{Zhao} G.-B.,  et~al., 2019, \mn@doi [\mnras] {10.1093/mnras/sty2845}, \href {https://ui.adsabs.harvard.edu/abs/2019MNRAS.482.3497Z} {482, 3497}

\bibitem[\protect\citeauthoryear{{Zhou}, {Zhang}  \& {Chen}}{{Zhou} et~al.}{2023}]{Zhou2023b}
{Zhou} S.,  {Zhang} P.,   {Chen} Z.,  2023, \mn@doi [\mnras] {10.1093/mnras/stad1824}, \href {https://ui.adsabs.harvard.edu/abs/2023MNRAS.523.5789Z} {523, 5789}

\makeatother
\end{thebibliography}

%%%%%%%%%%%%%%%%%%%%%%%%%%%%%%%%%%%%%%%%%%%%%%%%%%

% Don't change these lines
\bsp	% typesetting comment
\label{lastpage}
\end{document}